\documentclass[prl,aps,amssymb,showpacs,twocolumn]{revtex4-1}
\usepackage{amsmath}
\usepackage{amssymb}
\usepackage{amsthm}
\usepackage{amsfonts}
\usepackage{enumerate}
\usepackage{latexsym}
\usepackage{bm}
\usepackage{graphicx}
\usepackage{subfigure}

\newcommand{\beq}{\begin{equation}}
\newcommand{\eneq}{\end{equation}}

\newcommand{\bs}[1]{\boldsymbol{#1}}

\input{epsf}

\begin{document}

\tolerance 10000

\newcommand{\vk}{{\bf k}}

%\draft

\title{From density functional theory to the functional renormalization
  group: superconductivity in the iron pnictide LiFeAs}

\author{Christian Platt${}^1$}
\author{Ronny Thomale${}^2$}
%\author{B. Andrei Bernevig${}^2$}
\author{Werner Hanke${}^1$}

\affiliation{${}^1$Theoretical Physics, University of W\"urzburg, D-97074 W\"urzburg}
\affiliation{${}^2$Department of
Physics, Princeton University, Princeton, NJ 08544}

\begin{abstract}
  A combined density functional theory and functional renormalization
  group method is introduced which takes into account
  orbital-dependent interaction parameters to derive the effective
  low-energy theory of weakly to intermediately correlated Fermi
  systems. As an application, the competing fluctuations in LiFeAs are
  investigated, which is the main representative of the 111 class of
  iron pnictides displaying no magnetic order, but superconductivity,
  for the parent compound. The superconducting order parameter is
  found to be
  of $s_\pm$ type driven by collinear antiferromagnetic
  fluctuations. They eventually exceed the ferromagnetic fluctuations
  stemming from the small hole pocket at the $\Gamma$ point, as the
  system flows to low energies.
\end{abstract}

\date{\today}


\maketitle

%\section{Introduction}

In recent years the renormalization group (RG) has become a
much-used general concept to derive effective theories, e.g. at long
length scales or for a low-energy window of a given many-particle
system. For weakly to intermediately coupled fermion systems, one is
mainly interested in the effective interactions near the Fermi
surface $(E_\text{F})$, as they contain the relevant information about
possibly symmetry-broken (magnetic, superconducting, etc.) or
non-Fermi-liquid ground states: By systematically integrating out
''high-energy'' degrees of freedom one can, thus, access
competing orders at low-energy or temperature in the phase diagram.
The RG approaches to interacting fermions are less biased than
diagrammatic summations in a particular channel (such as e.g. the
RPA), as competing channels are treated on equal footing. To compute
the effective interactions near $E_\text{F}$, many recent works use the
RG flow equations for the effective action or one-particle
irreducible vertex functions~\cite{wetterich93plb90,salmhofer-01ptp1} that avoid some complications of
other straightforward adaptions of ''Wilsonian RG'' for interacting
fermions~\cite{shankar94rmp129}. These RG schemes are commonly named functional RG (FRG),
as they aim at keeping as much as possible of the wave vector and
frequency dependence of the vertex function, i.e. describe a flow of
a coupling function rather than a flow of a finite number of coupling
constants.
Due to their original motivation stemming from quantum-field theory
and statistical mechanics, the FRG approaches have so far been model
oriented: in the high-$T_c$ superconducting (SC) cuprates and
pnictides, for example, a variety of versions of effective Hubbard-type
models have been employed. They are usually  confined to a 
reduced Hilbert space as compared to the physical system and retain
only the short-range interactions as parameters.

In this work, we point out that the FRG, as a powerful scheme to
resolve competing orders, can be combined with
''a priori'' schemes such as the density functional theory (DFT).
This new feature allows us to introduce an unbiased connection
between electronic structure determinations and the competing ordering
tendencies in the phase diagram. 
%This is a general statement which is
%illustrated here, in an approximate manner, in the example of the
%recently much-discussed SC iron pnictides and their phase diagrams. 
We start at ''high-energy'' from a local density
(LDA)-type calculation. Through an evaluation of the Kohn-Sham
Hamiltonian~\cite{Dreizler1990} between maximally-localized Wannier functions (MLWF)~\cite{imada-10jpsj112001}, LDA determines the bare hopping parameters of the
model. Simultaneously, via a constrained screening of the direct and
exchange Coulomb interactions and calculation of the corresponding
MLWF matrix elements, it accounts for an ''a priori'' determination of
the interactions. Our proposal of a DFT-FRG method still relies on
some common approximations such as the use of frequency-independent effective
interactions (only the full $k$-dependence along the Fermi surface is
retained) and the neglect of the flow in the self energy, constraining
its applicability to electron systems in the weak to intermediate
coupling regime. For this case, however, it provides a precise
treatment from ab initio parameters to a low energy description.

%Nevertheless, the results
%obtained in this simplified LDA+FRG scheme demonstrate already a
%separation of ''what is more universal and what is more
%material-dependent'' in superconductivity in pnictides: the pairing
%mechanism appears magnetically (SDW) driven, i.e. via electronic
%correlations. On the other hand, when comparing two classes of
%pnictides (As-based versus P-based) material trends emerge: when
%replacing As by P, lower $T_c$-values, a propensity towards a nodal
%SC gap and the absence of a leading SDW (magnetic) instability are
%obtained, results which appear in qualitative accord with most
%experiments ().

In this Letter, we demonstrate the usefulness of the DFT-FRG formalism for the iron based SC
compound LiFeAs.  Soon after the synthesis of the 1111 and 122
pnictides~\cite{kamihara-08jacs3296}, LiFeAs as a representative of
the 111 family, has been detected, with a SC phase of $T_{\text{c}} \sim
16 \text{K}$~\cite{wang-08ssc538,tapp-08prb060505}.  One peculiar property of the 111 family is the absence
of magnetic order in the phase diagram, while an SC instability is
found already for the parent compound.  From the picture of itinerant
magnetism, this may be explained because of reduced nesting of the hole and electron
pockets in LiFeAs as opposed to the other pnictide families where
magnetic order is observed. This is suggested by band structure calculations~\cite{singh08prb094511} as well as by ARPES
measurements on LiFeAs~\cite{borisenko-10prl067002}. The latter are
particularly controlled due to the simple crystal structure (tetragonal
P4/nmm) and the nearly complete absence of surface states for
LiFeAs~\cite{lankau-cm1008}.

 \begin{widetext}
 \begin{figure*}[t!]
  \begin{minipage}[l]{0.76\linewidth}
    \includegraphics[width=\linewidth]{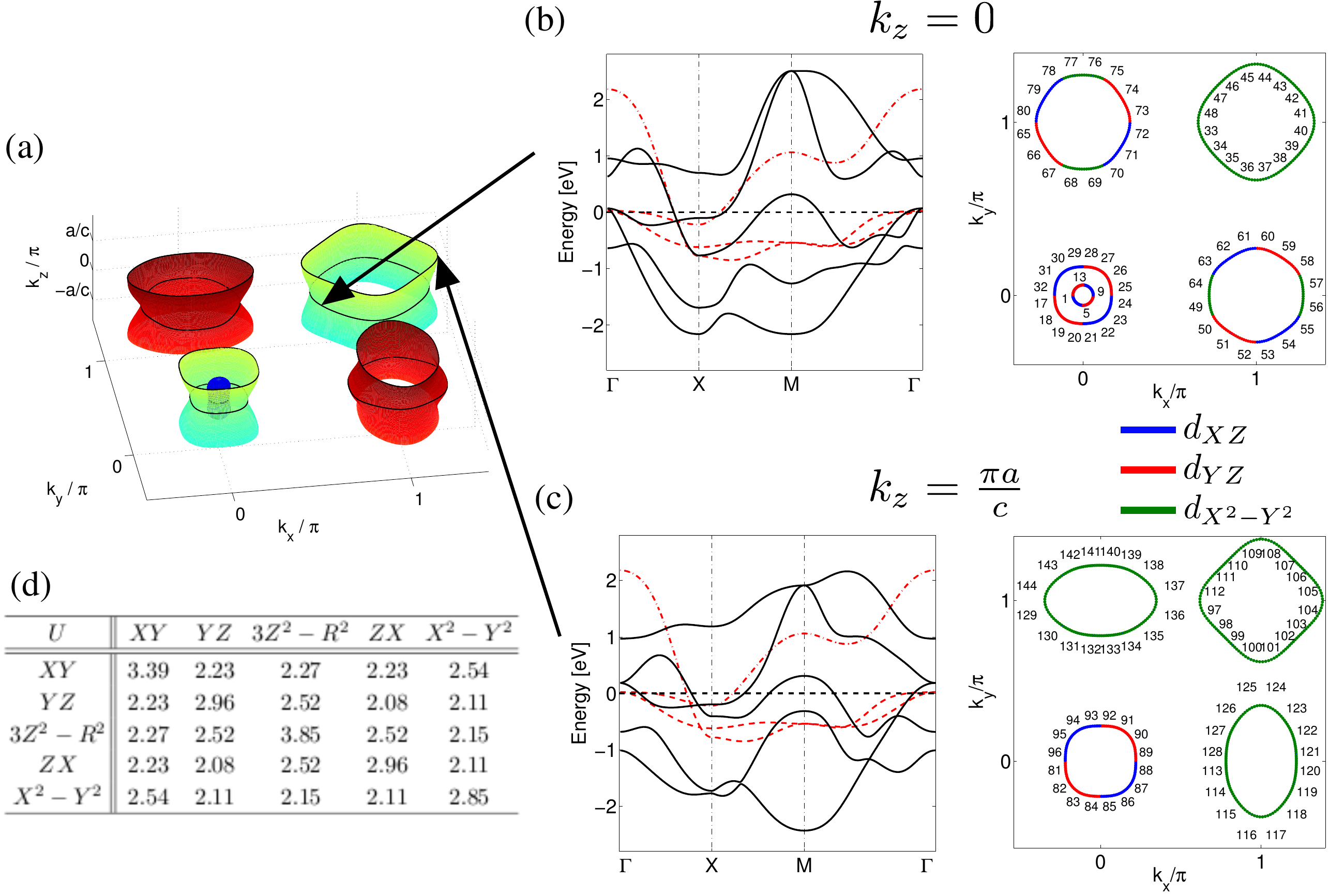}
  \end{minipage}
  \caption{(Color online). Summary of the ab-initio input on LiFeAS
    for the DFT-FRG. (a) 3d Fermi surface topology with visible dispersive
    features along $z$. 5-orbital band structure and Fermi surfaces
    for (b) $k_z=0$ cut of the bands (Brillouin zone patches from $1$ to $80$) and
    (c) $k_z=\pi$ cut (Brillouin zone patches from $81$ to $144$),
    corresponding to the patching in the insets of Fig.~\ref{pic2}. The
    red dashed lines denote the 3-band model used
    in~\cite{brydon-cm1009}. Brillouin zone patches and dominant
    orbital weights are indicated along the Fermi surfaces. (d)
    Orbital dependent interaction matrix $U_{ab}$ as obtained from
    cRPA calculations~\cite{miyake-10jpsj044705}. }
%The interaction
 %   scales vary significantly for different orbitals.}
    
 \label{pic1}
 \vspace{-0pt}
 \end{figure*}
 \end{widetext}

 Despite the absence of magnetic order in LiFeAs, the role and nature
 of the magnetic fluctuations are still key ingredients for the SC pairing. As the material seems to reside in the intermediately coupled regime
 and electron-electron interactions are still important, the driving
 mechanism of SC is expected to originate from
 magnetic fluctuations~\cite{phonon}.
%put this as a reference (there are also experimental findings which may
% be explained by phonon-mediated effects in this
% compound~\cite{gooch-09epl27005,kordyuk-cm1002}; still, the consensus
% is that the electronic interaction scale is dominant for
% LiFeAs). 
Knight shift experiments, as published up to now, provide indication for
 strong AFM fluctuations~\cite{jeglic-10prb140511}. ARPES studies, however, also point out the proximity of
 van Hove singularities to the Fermi level triggering
 ferromagnetic fluctuations~\cite{borisenko-10prl067002} which is
 further substantiated by transport measurements~\cite{heyer-cm1010}.  

A considerable body of experimental results point towards a (multi-)
 gap nodeless SC phase as given by NMR~\cite{inosov-10prl187001},
 specific heat and ARPES~\cite{sasmal-cm1004,stockert-cm1011} as well
 as microwave impedance and penetration depth
 measurements~\cite{imai-cm1009,song-cm1007}.  Many aspects of the
 SC phase, however, in particular its order parameter symmetry, are still under
 current debate.  While a major body of experimental evidence does not
 appear inconsistent with an anisotropic $s_\pm$ order parameter, the
 fishtail effect in LiFeAs, for example, is claimed to hint $p$-wave
 SC~\cite{pramanik-cm1009}. This has also been found by
 an RPA mean field study of a 3-band model used to represent the
 LiFeAs band structure~\cite{brydon-cm1009}. The results can be
 interpreted such that the small hole pocket at the $\Gamma$ point,
 being close to a  van-Hove singularity (due to the Stoner criterion) drives the
 ferromagnetic fluctuations, which in turn should trigger the
 formation of a $p$-wave instability. This picture derives from the  RPA
 and is consistent with a mean-field study where the ferromagnetic
 fluctuations are considered dominant. 
This exciting $p$-wave suggestion awaits still further confirmation
when (beyond the RPA) competing fluctuations are taken into account
and a more generic 5-orbital LiFeAs band structure as well as
orbital-dependent interactions are employed.
%independent of the Hubbard interaction
% scale $U$ up to comparably low temperatures~\cite{brydon-cm1009}. 
%It is unclear how
% strongly the deficiencies of the RPA approach itself, the big
% deviations from the 3-band model to the generic 5-orbital LiFeAs band
% structure (see Fig.~\ref{pic1}b and c), or the approximative choice
% of interactions (as opposed to fully orbital-dependent interactions
% as shown in Fig.~\ref{pic1}d) may spoil the final result.   
The competing magnetic fluctuations, as suggested by experimental evidence,
 necessitate a description which is as close to the experimental
 parameters as possible, taking full advantage of the high precision
 ab-initio electronic-structure data available. In what follows, the DFT-FRG approach takes into account the most accurate LDA data for the
 band structure and interactions. The FRG enables us to treat the different fluctuations of the system in
 an unbiased fashion (Fig.~\ref{pic1}).
%The existence of a gap in the superconducting phase is not yet fully
%settled
%DFT calculation: nesting reduced~\cite{singh08prb094511}
%negative pressure coefficient, high pressure limit of 1111 and 122
%compounds?~\cite{gooch-09epl27005}
%Knight shift: some indication for AFM fluctuations, but probably
%inconclusive~\cite{jeglic-10prb140511}
%ARPES: simple crystal structure (tetragonal P4/nmm) significant low-energy density of states, bad
%nesting, band renormalization similar to other
%pnictides~\cite{borisenko-10prl067002}, very clean results due to the
%nearly perfect absence of surface states~\cite{lankau-cm1008}
%life time measurements may suggest a role of phonon-mediated
%contributions in the superconducting phase, while the
%electron-electron is still assumed to be the dominant scale of the
%problem~\cite{kordyuk-cm1002}
%NMR: weakly coupled single gap
%superconductor~\cite{inosov-10prl187001}
%specific heat measurement hint multiple gaps~\cite{sasmal-cm1004},
%recent joint study with ARPES show particularly good agreement~\cite{stockert-cm1011}
%penetration depth hint two gaps with decreased gaps away from optimal doping~\cite{song-cm1007}
%RPA / mean field prediction from a 3 band model: $p$-wave superconductivity~\cite{brydon-cm1009}
%gap: microwave impedance measurements~\cite{imai-cm1009}
%fishtail effect: similar to $p$-wave
%superconductor~\cite{pramanik-cm1009}

 We start with the construction of an ab-initio effective
 Hamiltonian for our compound. As an input to our FRG
 calculations, we employ data from a recent work by Miyake et
 al., where details of the ab-initio
 procedure can be found~\cite{miyake-10jpsj044705}. The first step is a conventional
 band-structure calculation in the framework of LDA. From there, ''target
 bands'' are chosen around $E_\text{F}$, which define the band complex and
 the corresponding orbitals (in our case the five d-orbitals of the Fe-3d
 electrons). Simultaneously, the  MLWF's are extracted,
 which, via their matrix elements of the Kohn-Sham Hamiltonian
 $H_{KS}$, determine the transfer integrals $t_{mn}(\vec{R})$
\begin{equation}
t_{mn}(\vec{R}) = \langle \phi_{m\vec{0}}|H_{KS}|\phi_{n\vec{R}}\rangle.
\end{equation}
Here $\phi_{n\vec{R}}(\vec{r})$ denotes the MLWF centered at site
$\vec{R}$ for the $n$-th orbital. The one-body Hamiltonian part
 is
\begin{equation}
H_0=\sum_{i,j,\sigma}\sum_{m,n}t_{mn}(\vec{R}_i - \vec{R}_j){a^{\sigma}_{in}}^{\dagger}{a^{\sigma}_{in}}^{\phantom{\dagger}},
\end{equation}
where ${a^{\sigma}_{in}}^{\dagger}$
$({a^{\sigma}_{in}}^{\phantom{\dagger}})$ are the creation
(annihilation) operators of the MLWF with spin $\sigma$. In the second
step, effective interaction parameters are extracted in terms of
MLWF-matrix elements~\cite{miyake-10jpsj044705}. A partially screened Coulomb
interaction at zero frequency $W(\vec{r},\vec{r'};\omega=0)$ is
calculated in the so-called cRPA, i.e. with the constraint that for
the "high-energy" non-target bands RPA screening is employed~\cite{imada-10jpsj112001}. 
%screening channels inside the target band are ''cut out'' [CITE]. 
Note, that the
Coulomb interactions $U_{mn}(\vec{R}) = \langle \phi_{m\vec{0}}
\phi_{m\vec{0}}|W|\phi_{n\vec{R}}\phi_{n\vec{R}}\rangle $ and exchange
interactions $J_{mn}(\vec{R}) = \langle \phi_{m\vec{0}}
\phi_{n\vec{0}}|W|\phi_{n\vec{R}}\phi_{m\vec{R}}\rangle $ are orbital
dependent, comprising the interaction Hamiltonian:
\begin{align}
H_I &= \frac{1}{2}\sum_{\sigma\delta}\sum_{ij}\sum_{nm}{\Big\{}U_{mn}(\vec{R}_i - \vec{R}_j)a^{\sigma\dagger}_{in}a^{\rho\dagger}_{jm}a^{\rho}_{jm}a^{\sigma}_{in} \\\nonumber
&+ J_{mn}(\vec{R}_i - \vec{R}_j)\left(a^{\sigma\dagger}_{in}a^{\rho\dagger}_{jm}a^{\rho}_{in}a^{\sigma}_{jm}
+ a^{\sigma\dagger}_{in}a^{\rho\dagger}_{in}a^{\rho}_{jm}a^{\sigma}_{jm}\right){\Big\}}.
\end{align}
%Here, two comments are in order: 
One challenging problem of the ''down-folding'' is that, for entangled
bands, it is not clear a priori how to ''cut out'' the d-subspace of
the Fe-orbitals, and how to unambiguously distinguish the screening
channels within the d-space from the total screening. (For a practical
solution of this problem see again~\cite{imada-10jpsj112001} and
references therein.) 

%Miyake et. al.
%have recently developed a procedure to overcome this difficulty and
%successfully applied it to transition metals [CITE].

%The second point
%concerns the static-screening approximation in $U$ and $J$. FRG
%calculations with an energy-dependent interaction can, in principle,
%be performed, but this becomes exceedingly difficult. On the other
%hand, recent work [] has qualitatively discussed how an appropriate
%model with a static $U$ can be constructed, employing a modified
%self energy. While the self energy near $E_\text{F}$ is of importance it is
%harder to obtain in the FRG []. Moreover, many trends in the
%self energy, like gap openings, are ''fore-shadowed'' in the
%effective interactions, the flow of which to low energies we
%consider here. Thus, we relegate the $\omega$-dependence of the
%effective interactions $U$ and $J$ and the self energy calculations to a
%forth-coming study.

\begin{figure}[t]
  \begin{minipage}[l]{0.85\linewidth}
    \includegraphics[width=\linewidth]{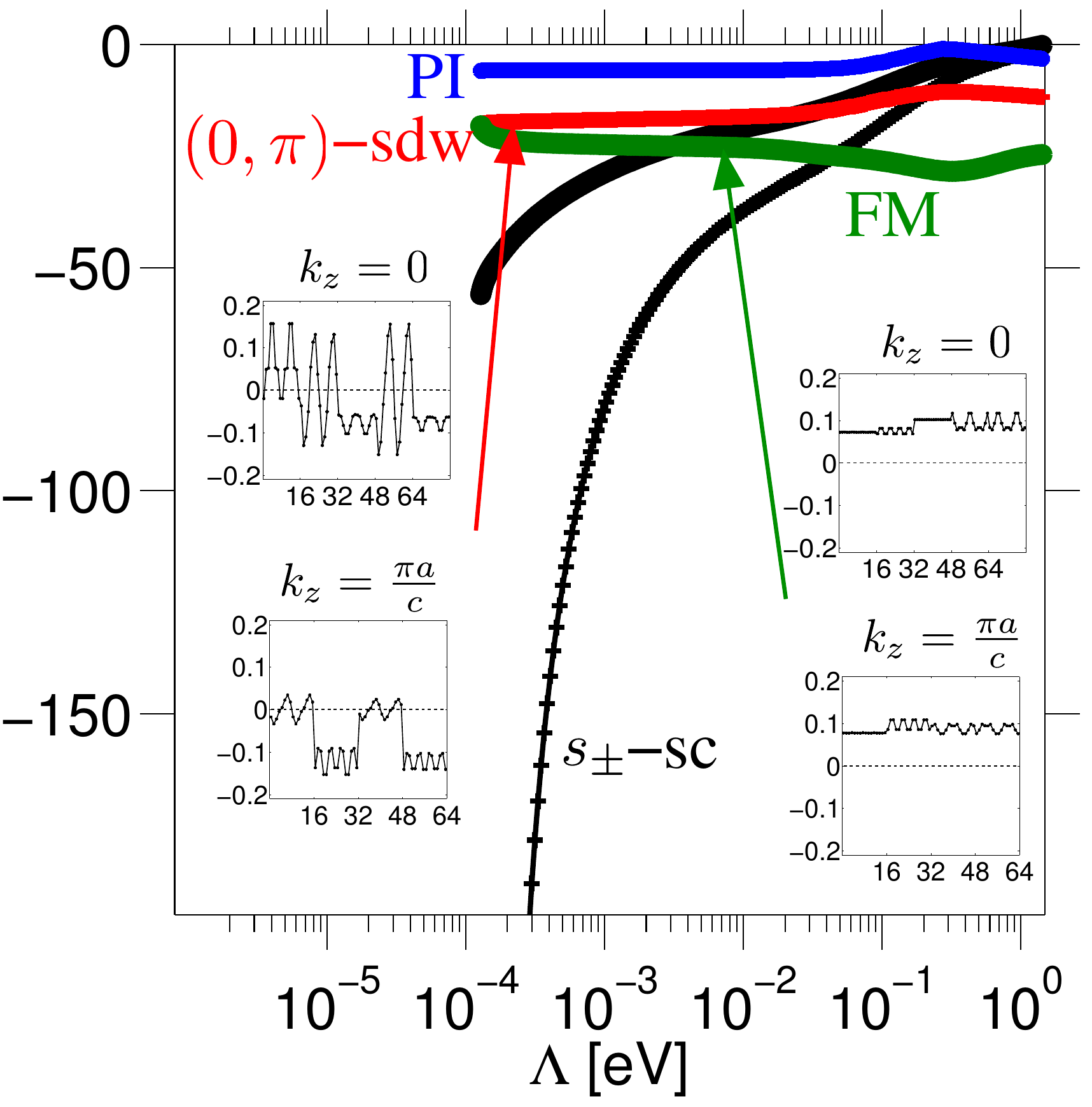}
  \end{minipage}
  \caption{(Color online). FRG temperature flow for LiFeAs. The form factors for
    nodal SDW and FM fluctuations are shown in the left and right
    inset. At cutoff scales at $\Lambda \sim 0.01 eV$ the SDW
    becomes competitive to the previously dominant FM. The
    leading instability is of SC $s_\pm$ type with the form factor
    displayed in Fig.~\ref{pic3}.}
\label{pic2}
\vspace{-0pt}
\end{figure}

The final data of the ab-initio calculations consisting of band
structure, Fermi surfaces, and bare interaction parameters are summarized in
Fig.~\ref{pic1}. The 111 compounds have a relevant dispersion
orthogonal to the FeAs layers
(Fig.~\ref{pic1}a) for which we consider both the $k_z=0$ and
$k_z=\pi$ cut. The main features of the $z$-dispersion are the change
of orbital weight along the Fermi surface as well as the absence of
the small second hole pocket at $\Gamma$ for $k_z=\pi$. As seen from
the interaction matrix, e.g. the intra-orbital interactions (with
comparably high absolute interaction scales up to $3.85$ eV) differ by
more than $30\%$ between different orbitals, stressing the need to
consider fully orbital-dependent parameters.

%TO DO BAND STRUCTURE PLUS ORBITAL INTERACTION MATRIX
%comment on that absolute interaction scale is higher (high pressure
%limit of 122!) check back with Fig.~\ref{pic1}, no single equation for
%band structure anymore.

\begin{figure}[t]
  \begin{minipage}[l]{0.99\linewidth}
    \includegraphics[width=\linewidth]{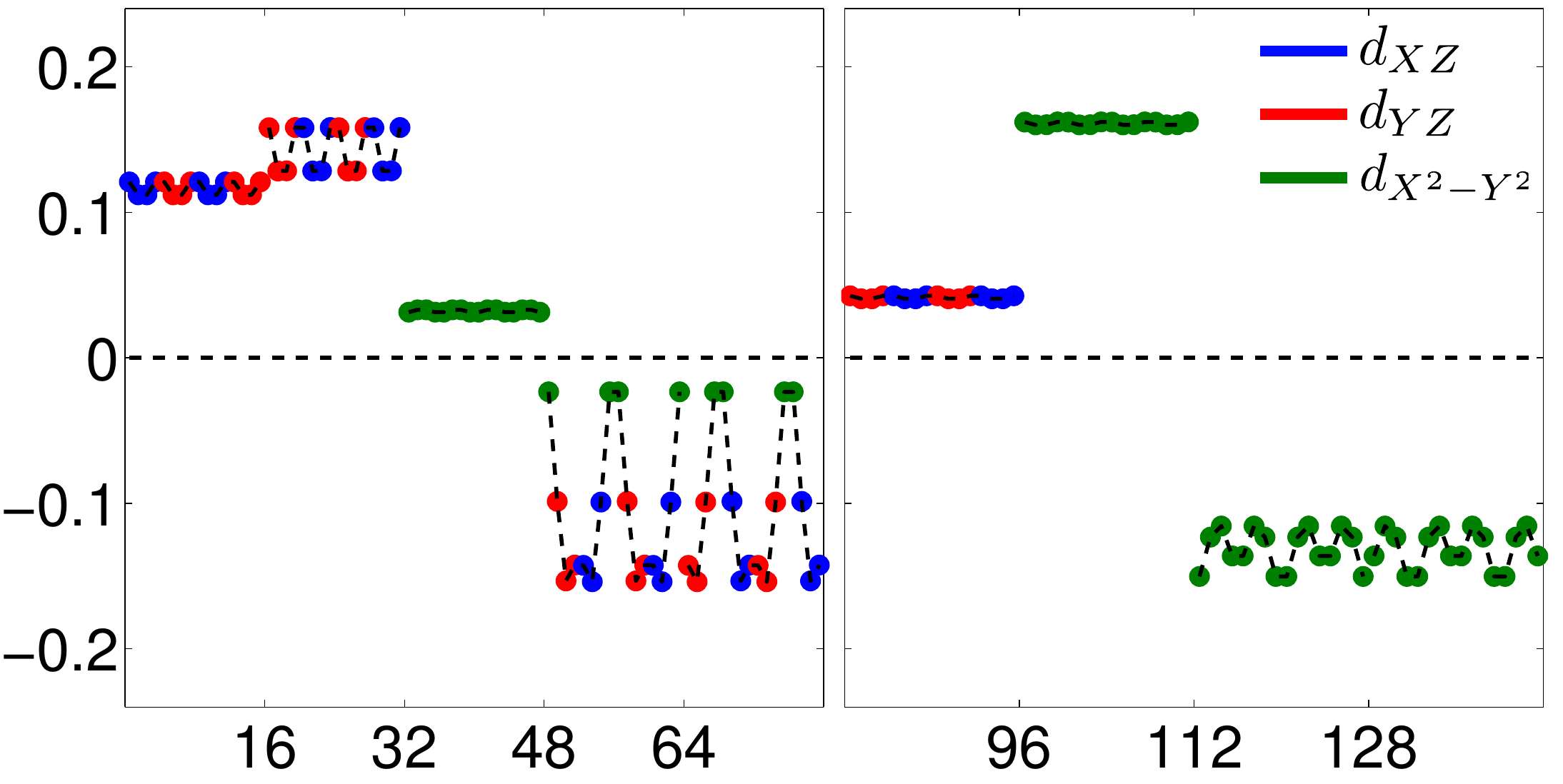}
  \end{minipage}
  \caption{(Color online). Leading SC form factor along the cuts for $k_z=0$
  and $k_z=\pi$. The dominant orbital weights for the different
  patches are indicated by color. The form factor shows a multi-gap
  nodeless anisotropic $s_\pm$ order parameter.}
\label{pic3}
\vspace{-0pt}
\end{figure}

Using multi-band
FRG~\cite{thomale-09prb180505,wang-09prl047005,thomaleasvsp,platt-09njp055058,wang-10prb184512},
high-energy electronic excitations $(\varepsilon > \Lambda)$ are
recursively integrated out, arriving at an effective low-energy
interaction or 4-point vertex function (4PF)
$V(\bs{k_1}n_1,\bs{k_2}n_2, \bs{k_3}n_3, \bs{k_4}n_4)$. Here,
$n_1,\ldots n_4$ label the different (in our case five) bands of the
band complex considered and $\bs{k_1}$ to $\bs{k_4}$ the incoming and
outgoing momenta. When the infrared cutoff $\Lambda$ approaches the
Fermi surface, a diverging renormalized 4PF then signals a
corresponding instability towards a symmetry-broken phase, with
$\Lambda_c$ serving as an upper bound for the transition temperature
$T_c$.  In addition, we also employ a temperature-flow FRG and
compare it to the results obtained by the conventional cutoff-flow FRG~\cite{honerkamp-01prb184516}. At each renormalization iteration, one sums over the five
one-loop
diagrams~\cite{honerkamp-01prb035109,halboth-00prb7364,zanchi-00prb13609},
i.e. over the Cooper, spin-density wave (SDW), screening and
vertex-correction channels to arrive at the renormalized vertex
function. 
%In doing so, one takes into account the most important
%fluctuations on equal footing for each differential step in
%$\Lambda$. 
Technical details of our FRG procedure can be found
elsewhere~\cite{thomale-09prb180505,wang-09prl047005,thomaleasvsp}. Due
to practical limitations, approximations are made, such as the neglect
of the frequency dependence of the 4-point vertex function
$V_{\Lambda}$ and projecting the external momenta
$\bs{k_1},\ldots\bs{k_3}$ onto the Fermi surface (with $\bs{k_4}$
being determined from momentum conservation).  For a given instability
characterized by some order parameter $\hat{O}_{\bs{k}}$ (the most
important example of which is the SC instability
$\hat{O}^{SC}_{\bs{k}} = c_k c_{-k}$ for LiFeAs), the 4PF in the
particular ordering channel can be written as
$\Sigma_{k,p}V_\Lambda(k,p)[\hat{O}^{\dagger}_{\bs k}\hat{O}_{\bs p}]$.
 It can be decomposed into different eigenmode contributions
$V_{\Lambda}^{SC}(\bs{k},-\bs{k},\bs{p}) = \sum_i
c_i^{SC}(\Lambda)f^{SC,i}(\bs{k}) f^{SC,i}(\bs{p})$ where $i$ is an
enumeration index. The leading instability of a given channel
corresponds to an eigenvalue $c^{SC}_1(\Lambda)$ first
diverging under the flow of $\Lambda$.  $f^{SC,1}(k)$ is the SC form
factor of the leading SC pairing mode which tells us about the pairing
symmetry and, hence, gap structure. For all different
ordering channels, the form factors are computed along the discretized
Fermi surfaces. (Shown are the ferromagnetic (FM) and spin density wave
(SDW) form factor in the insets of Fig.~\ref{pic2} as well as the SC
form factor in Fig.~\ref{pic3}.)

%MISSING TO DO: Understand the ferromagnetic form factor as is and plot
%it in Fig.~\ref{pic2}.
The flow of these leading eigenvalues of the instability channels is summarized in Fig.~\ref{pic2}. We find that, for high cutoffs
(temperatures), the FM fluctuations are dominant in correspondence to
previous studies~\cite{brydon-cm1009}. By removing the small hole
pocket at $\Gamma$, we checked that it is, indeed, the main resource for FM
fluctuations. At intermediate scales, however, we see in
Fig.~\ref{pic2} that the
collinear SDW fluctuations, driven by hole to electron pocket
scattering along $(\pi,0)$ and $(0,\pi)$, become competitive and finally
seed an $s_\pm$ instability. (This result is identically
obtained both for the cutoff and the temperature flow parameter
formulation. The latter is more adequate to track FM fluctuations.) From there, the vertex flow
can be understood along the lines of other pnictide families~\cite{wang-09prl047005,thomaleasvsp}. 
%In particular, the nodal SDW fluctuations
%provide an alternative explanation for the high low energy density of
%states observed in LiFeAs which need not necessarily be due to nodal
%SC.

The form factor of the SC instability for LiFeAs is plotted in
Fig.~\ref{pic3} along the $k_z=0$ and $k_z=\pi$ cuts. The $k_z=0$ cut
looks typical as for other pnictide families and displays an $s_\pm$
form factor changing sign between hole and electron pockets with
strong anisotropy features along the electron pockets. As indicated by
the dominant orbital weights along the different patches, we find from
Fig.~\ref{pic3} that the form factor can be well understood in a
scenario where the SC form factor seeks to minimize intra-orbital
repulsion which is the most relevant interaction scale~\cite{platt-cm1012}. For
$k_z=\pi$, the largest hole-pocket gap is given by the $M$ pocket as
it can scatter to the electron pockets at $X$ through intra-orbital
interactions. Since the orbital weight for these pockets is uniformly
of $d_{X^2-Y^2}$-type, the form factor anisotropies on the electron pockets are rather small.

%TO DO: Incorporate comments on Plot 1: Band structure (including 3-band approximation), include
%kz0 and kzpi, orbital weights, comment on nesting. Main idea: kz=0:
%small hole pocket at Gamma present, contributed ferromagnetic
%fluctuation which die out towards lower energies as SDW fluctuation
%channel becomes more dominant (in particular for kz=pi).  kz=pi: M
%pocket better nested with electron pockets and fits with orbital
%weight. probably even double column, showing also the 3d zylinder
%dispersion? Maybe put the band interaction matrix also here
%Plot 2: Flow of eigenvalues: discuss how the FM fluctuations die out
%for lower cutoff. IMPORTANT: Are the FM fluctuations clearly higher
%for kz=0 than for kz=pi? Single column pointers to where FM
%fluctuations die out (are they similar for kz=0 and pi?)
%Plot 3: Form factor of leading SC instability: comment on multi gap
%scenario, arrows for decisive scattering channel, single column
% 

In summary, we have introduced the DFT-FRG formalism to describe
Fermi surface instabilities by taking into account a-priori DFT (LDA)
band parameters and orbital-dependent interactions and providing a low-energy description in an unbiased
fashion. We have applied the DFT-FRG to LiFeAs and find the leading SC
instability to be of $s_\pm$-type. Within FRG, the FM fluctuations at
"high temperatures" are overcome by SDW fluctuations in the
effective low-energy sector.

%Jeroen van den Brink: mass renormalization underestimated by ab initio band
%calculation - effective mass deviation up to factor of 3 - so bands
%are probably slightly flatter in the fermi level vicinity and may give
%a very small bias against the strength of flat band induced
%ferromagnetic fluctuations. Still we find it to be very stable that p
%wave is overcome by pi 0 fluctuations.

%christian hess: quasiparticle interference seems to suggest p-wave,
%but only given that you take isotropic extended s-wave. However, the
%correspondence will for sure be better when anisotropic s-wave is assumed.

\begin{acknowledgments}
  We thank A. Chubukov, S.~Graser, M.~Daghofer, I.~Eremin, P.~Hirschfeld,
  D.~Scalapino, and J. van den Brink for discussions. The
  work was supported by DFG-SPP 1458/1 and by the
  Bavarian KONWIHR Program. RT is supported by the Humboldt Foundation.
%  and by Sloan Foundation.  
 % BAB is
%  supported by Princeton Startup Funds, Alfred P. Sloan Foundation,
%  NSF DMR- 095242, NSF China 11050110420, and MRSEC grant at Princeton
%  University, NSF DMR-0819860.
\end{acknowledgments}

%\begin{thebibliography}{11}
%%\bibitem{raghu} S. Raghu et al., Phys.Rev.B {\bf 77}, 220503 (2008)
%\end{thebibliography}

%\bibliographystyle{prsty}
%\bibliography{/users/tkm/rachel/bib/paper,/users/tkm/rachel/bib/book,/users/tkm/rachel/bib/htc,/users/tkm/rachel/bib/martin,/users/tkm/rachel/bib/na,/users/tkm/rachel/bib/unpub.bib}
%\bibliography{paper,book}

\begin{thebibliography}{10}
\bibitem{phonon}
There are also experimental findings which may be explained by phonon-mediated
  effects in LiFeAs~\cite{gooch-09epl27005,kordyuk-cm1002}; still, the
  consensus is that the electronic interaction scale is dominant.
\end{thebibliography}

\end{document}